\documentclass[aip,jcp,superscriptaddress,amsmath,amssymb,twocolumn,reprint]{revtex4-1}

\usepackage[version=3]{mhchem} 
\usepackage{graphicx}
\usepackage{amsmath}
\usepackage{amssymb}
\usepackage{nicefrac}
\usepackage{booktabs}

\usepackage{array}
\newcolumntype{m}{>{$} c <{$}}
\usepackage{color}

\def\rv{{\bf r}}
\def\fv{{\bf f}}

\def\beq{\begin{equation}}
\def\eeq{\end{equation}}

\def\mod{{\rm{MRF}}}

\begin{document}     

\author{Stefan Vuckovic}
\affiliation{Department of Theoretical Chemistry and Amsterdam Center for Multiscale Modeling, FEW, Vrije Universiteit, De Boelelaan 1083, 1081HV Amsterdam, The Netherlands}
\author{Paola Gori-Giorgi}
\email{p.gorigiorgi@vu.nl}
\affiliation{Department of Theoretical Chemistry and Amsterdam Center for Multiscale Modeling, FEW, Vrije Universiteit, De Boelelaan 1083, 1081HV Amsterdam, The Netherlands}

\title{Simple fully non-local density functionals for the electronic repulsion energy}
\begin{abstract}
From a simplified version of the mathematical structure of the strong coupling limit of the exact exchange-correlation functional, we construct an approximation for the electronic repulsion energy at physical coupling strength, which is fully non-local. This functional is self-interaction free and yields energy densities within the definition of the electrostatic potential of the exchange-correlation hole that are locally accurate and have the correct asymptotic behavior. The model is able to capture strong correlation effects that arise from chemical bond dissociation, without relying on error cancellation. These features, which are usually missed by standard DFT functionals, are captured by the highly nonlocal structure, which goes beyond the {\it Jacob's ladder} framework for functional construction, by using integrals of the density as the key ingredient. Possible routes for obtaining the full exchange-correlation functional by recovering the missing kinetic component of the correlation energy are also implemented and discussed. 
\end{abstract}

\maketitle
The widespread success of Kohn--Sham density-functional theory (KS DFT)\cite{KohSha-PR-65,CohMorYan-CR-12,Bur-JCP-12,Bec-JCP-14} across various chemical and physical disciplines has been also accompanied by spectacular failures,\cite{CohMorYan-CR-12} reflecting fundamental issues in the present density functional approximations (DFAs) for the exchange--correlation (XC) functional. Well-known examples are the paradigmatic case of the dissociation curves of the H$_2$ and H$_2^+$ molecules.\cite{Sav-INC-96,CohMorYan-CR-12} 
The usual DFAs approach to construct XC functionals consists in making an ansatz in terms of ``Jacob's ladder'' ingredients: \cite{PerRuzTaoStaScuCso-JCP-05,Bec-JCP-14,MedBusJiaPer-SCI-17,Ham-SCI-17} the local density, its gradient, its laplacian and/or KS kinetic energy density, up to occupied and virtual KS orbitals. While this strategy has been very successful for moderately correlated systems (see, e.g., Refs.~\onlinecite{Bur-JCP-12,Bec-JCP-14,ZhaTru-ACR-08,SunRemZhaSunRuzPenYanZenPauWagWuKlePer-NATC-16,ErhBleGor-PRL-16}), it has failed so far when correlation effects become important  (e.g., in stretched bonds, but also at equilibrium gemoetries when partially filled $d$ and $f$ subshells are present). This fact suggests that a different approach to DFAs is needed to address the problem of strong correlation.\cite{Bur-JCP-12,CohMorYan-CR-12,MorCoh-PCCP-14,CohMorYan-SCI-08,locpaper} 

The strong-interaction limit of DFT\cite{Sei-PRA-99,SeiGorSav-PRA-07,GorVigSei-JCTC-09,ButDepGor-PRA-12} provides information on how the exact XC functional depends on the density in a well defined mathematical limit, which is relevant for strong correlation. The thorough explorations of this limit reveal a mathematical structure totally different from that of Jacob's ladder ingredients. Instead of the local density, density derivatives or KS orbitals, in this limit we see that certain {\em integrals} of the density play a crucial role, encoding highly nonlocal information, \cite{Sei-PRA-99,SeiGorSav-PRA-07,GorVigSei-JCTC-09}
embodied in the so-called \textit{strictly-correlated electrons} (SCE) functional.\cite{Sei-PRA-99,SeiGorSav-PRA-07,GorVigSei-JCTC-09} This functional appears to be well--equipped for solving long-standing DFAs problems: it is self--interaction free, it captures the physics of charge localization due to strong correlation without resorting to symmetry breaking,\cite{MalMirCreReiGor-PRB-13,MenMalGor-PRB-14,MalMirMenBjeKarReiGor-PRL-15} and its functional derivative displays (in the low-density asymptotic limit) a discontinuity on the onset of fractional particle number.\cite{MirSeiGor-PRL-13} Despite these appealing features, there are two main obstacles to the routine use of the SCE functional: its availability is restricted to small systems\cite{SeiGorSav-PRA-07,VucWagMirGor-JCTC-15} and its energies are way too low for most of physical and chemical systems.\cite{MalMirCreReiGor-PRB-13,MalMirGieWagGor-PCCP-14,CheFriMen-JCTC-14,VucWagMirGor-JCTC-15} The \textit{nonlocal radius} (NLR) functional,\cite{WagGor-PRA-14} and the newer shell--model\cite{BahZhoErn-JCP-16} are inspired to the SCE functional form and retain only some of its non-locality. They are readily available,\cite{BahZhoErn-JCP-16} but, being approximations to the SCE functional, their energies are also too low with respect to those of chemical systems.\cite{WagGor-PRA-14,BahZhoErn-JCP-16} The information encoded in the SCE functional or its approximations can be combined with the complementary information from the weak coupling limit. This has been recently used for constructing XC functionals from a local interpolation along the adiabatic connection.\cite{locpaper,ZhoBahErn-JCP-15,BahZhoErn-JCP-16,VucIroWagTeaGor-PCCP-17} Although this approach is promising for treating strong correlation within the realm of DFT,\cite{locpaper} it can still easily over-correlate (for example for stretched bonds it overcorrelates the fragments), again because the SCE (exact or approximate) quantities are often far from the physical ones.\cite{VucIroWagTeaGor-PCCP-17} 

Nonetheless, the way in which the information encoded in the density is transfomed into an electron-electron repulsion energy in the SCE functional is very intriguing, with many physical appealing features.\cite{SeiGorSav-PRA-07,MalGor-PRL-12,LanDiMGerLeeGor-PCCP-16,SeiDiMGerNenGieGor-PRA-17} Motivated by this observation, in this letter we use the SCE mathematical structure to devise a new way to design fully non-local approximate density functionals for the electronic interaction energy at the physical coupling strength. Capturing the main structural motives of the SCE functional, we preserve many of its appealing features, but with repulsion energies that are much closer to those of physical systems. Moreover, besides accurate total repulsion energies, our model provides energy densities within the definition of the electrostratic potential of the XC hole that are also {\it locally} very close to exact ones, making it an ideal tool for the developement of functionals that use the exact exchange energy density, like hyperGGA's\cite{PerSmi-INC-01,Bec-JCP-05,PerStaTaoScu-PRA-08,KonPro-JCTC-15} or local hybrids.\cite{JarScuErn-JCP-03,ArbKau-CPL-07} In other words, it is known\cite{CleCha-JCP-93,Bec-JCP-93a,PerSmi-INC-01,PerStaTaoScu-PRA-08} that in order to use the exact exchange energy density we need a fully non-local correlation functional compatible with it. It is the purpose of this work to provide a new strategy to build this fully non-local functional at a computational cost similar to the one of the exact exchange energy density. 

In order to explain our construction we have to first quickly review some basic DFT equations. An exact expression for the XC energy can be obtained from the density-fixed adiabatic connection formalism (AC):\cite{LanPer-SSC-75,GunLun-PRB-76}
\begin{equation}
E_{xc}[\rho]= \int_0^1 W_\lambda[\rho]\mathrm{d}\lambda,
		\label{eq:ac_xc}
\end{equation}
where $W_\lambda[\rho]$ is the global (i.e., integrated over all space) AC integrand:
\begin{equation}
W_\lambda[\rho]= \langle \Psi_{\lambda}[\rho]  | \hat{V}_{ee}| \Psi_{\lambda}[\rho]\rangle - U[\rho].
		\label{eq:10.w_lam}
\end{equation}
The wavefunction $\Psi_{\lambda}[\rho]$ depends on the positive coupling constant $\lambda$ and minimizes $\langle \hat{T} +\lambda \hat{V}_{ee} \rangle$, while integrating to $\rho(\rv)$, the density of the physical system ($\lambda=1$). This way, AC links the KS non--interacting state described by $\Psi_0[\rho]$ and the physical state described by $\Psi_1[\rho]$. It also further connects the physical and the SCE state, i.e. the state of perfect electron correlation, corresponding to the limit $\lambda\to\infty$. The XC energy densities along the adiabatic connection ({\it i.e.}, position-dependent quantities $w_\lambda(\rv)$ that integrate to $W_\lambda[\rho]$ when multiplied by the density) are not uniquely defined and therefore we have to be specific on their {\it gauge}.\cite{BurCruLam-JCP-88,CruLamBur-JPCA-98,TaoStaScuPer-PRA-08,locpaper,VucIroWagTeaGor-PCCP-17} A physically sound gauge often considered in DFT is the one of the electrostatic potential of the XC hole.\cite{Bec-JCP-05,BecJoh-JCP-07,PerStaTaoScu-PRA-08,locpaper} Within this gauge we can express the $\lambda$-dependent energy density in terms of the corresponding spherically-averaged XC hole\cite{Bec-JCP-05,PerStaTaoScu-PRA-08,GorAngSav-CJC-09,locpaper} $h_{\rm xc}^\lambda(\rv,u)$ obtained from $|\Psi_{\lambda}[\rho]|^2$, 
\beq \label{eq:wxc_def}
    w_\lambda(\rv) = \frac{1}{2} \int_0^\infty \frac{h_{\rm xc}^\lambda(\rv,u)}{u}4 \pi u^2 \mathrm{d}u,  
\eeq
where $u= \left | \rv-\rv' \right|$ is the distance from a reference electron in $\rv$.  In the SCE ($\lambda\to\infty$) limit the energy density $w_\lambda(\rv)$ in the gauge of Eq.~\eqref{eq:wxc_def} has the exact form\cite{MirSeiGor-JCTC-12}
\beq \label{eq:wxcSCE}
w_\infty(\rv)=\frac{1}{2}\sum_{i=2}^N\frac{1}{|\rv-\fv_i([\rho];\rv)|}-\frac{1}{2}v_H(\rv),
\eeq
where $v_H(\rv)$ is the Hartree potential and the {\em co-motion functions} $\fv_i([\rho];\rv)$ are non-local functionals of the density that give the positions of the remaining $N-1$ electrons when one electron is at position $\rv$.\cite{Sei-PRA-99,SeiGorSav-PRA-07,SeiDiMGerNenGieGor-PRA-17} From Eq.~\eqref{eq:wxcSCE} we see that in the $\lambda\to\infty$ limit the energy density $w_\infty(\rv)$ is fully determined by the distances $R_i^{\rm SCE}([\rho];\rv)=|\rv-\fv_i([\rho];\rv)|$ between a reference electron in $\rv$ and the remaining $N-1$ ones. For example, in the case of one-dimensional systems, the distances $R_i^{\rm SCE}([\rho];x)$ can be constructed exactly\cite{Sei-PRA-99,ColDepDim-CJM-15} from the equations (with $i=2,\dots,N$)
\beq
\label{eq:radii1D}
\int_x^{f_i(x)}\rho(x')\, {\rm d} x'=i-1, \qquad R_i^{\rm SCE}(x)=|x-f_i(x)|,
\eeq
which can be solved in terms of the function $N_{\rm 1D}(x)=\int_{-\infty}^x\rho(x')\, {\rm d} x'$ and its inverse $N_{\rm 1D}^{-1}(y)$.\cite{Sei-PRA-99,MalGor-PRL-12,MalMirCreReiGor-PRB-13} We see that in this limit each electron is separated by the closest one by a piece of density that integrates exactly to 1 (in other words, fluctuations are totally suppressed in the limit of extreme correlation), with the key ingredient being the amount of expected electrons between two electronic positions. 

In this work we propose a way to generalize the SCE form of Eq.~\eqref{eq:wxcSCE} by using $\lambda$-dependent distances (or ``radii'') $R_i^{\lambda}([\rho];\rv)$ that will take into account the effect of fluctuations, which are not as suppressed as in the extreme SCE case. Thus, our ``multiple-radii functional'' ($\mod$) energy density reads as 
\beq
w_\lambda^\mod(\rv)= \frac{1}{2} \sum_{i=2}^{N} \frac{1}{R_i^{\lambda}([\rho];\rv)}-\frac{1}{2}v_H(\rv).
\label{eq:edens_mod}
\eeq
As we shall see, we will determine the $R_i^{\lambda}([\rho];\rv)$ by using a simplified version of the same kind of integrals of the density that appear in the SCE limit, introducing the average effect of fluctuations by reducing the amount of expected charge between two electronic positions. Before coming to the details of the  $R_i^{\lambda}([\rho];\rv)$ construction, we remark that Eq.~\eqref{eq:edens_mod} can be also derived from the
following model for the spherically-averaged pair-density
\beq
P_{2,\lambda}^{\mod}([\rho];\rv,u)=\frac{1}{4\pi u^2}\sum_{i=2}^{N}\rho(\rv)\delta\big(u-R_i^\lambda(\rv)\big),
\label{eq:p2}
\eeq
where $\delta$ is the Dirac delta function. Given that the model of Eq.~\eqref{eq:p2} is properly normalised, the corresponding $\lambda$-dependent XC hole satisfies the sum rule, integrating to $-1$ electron.
\begin{figure}
\includegraphics[width=8.5cm]{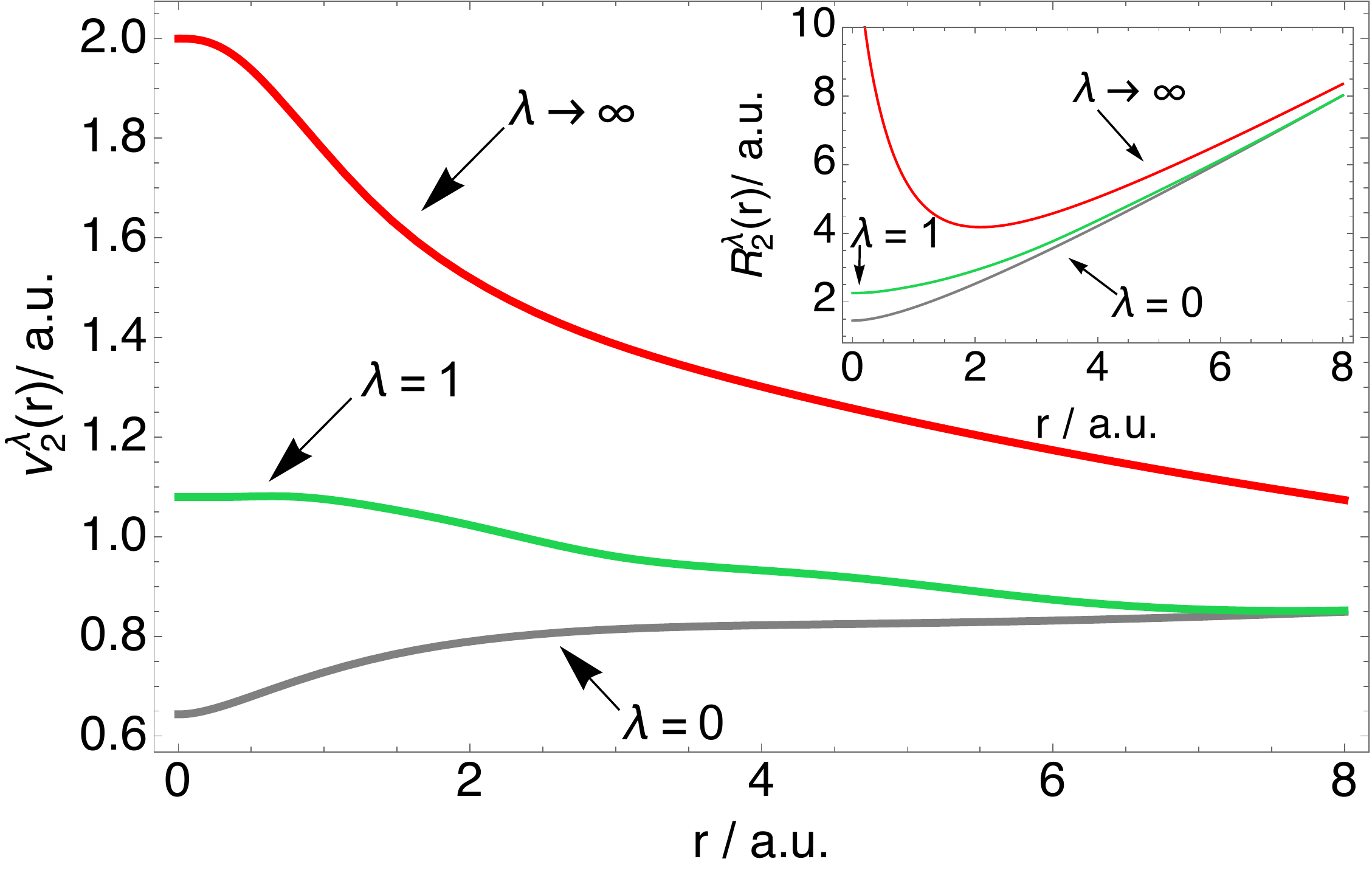}
\caption{The $\nu_2^\lambda(\rv)$ quantity of Eq~\eqref{eq:nu} and the corresponding $R_2^{\lambda}(\rv)$ radius at different coupling strengths for the hydride ion obtained from Eq.~\eqref{eq:rad_inver} using accurate xc energy densities $w_\lambda(\rv)$ from Refs.~\onlinecite{IroTea-MP-15,locpaper}.}
\label{fig_invrad}
\end{figure} 

We now turn to the construction the radii $R_i^{\lambda}([\rho];\rv)$.  Motivated by the structure of Eq.~\eqref{eq:radii1D},
and similarly to the recent non-local approximations for the SCE functional for three-dimensional systems,\cite{WagGor-PRA-14,BahZhoErn-JCP-16} we introduce the spherically averaged density $\tilde{\rho}(\rv,u)$ around a position $\rv$,
\begin{equation}\label{eq:spher_rho}
\tilde{\rho}(\rv,u)=\int \frac{1}{4 \pi} \rho(\rv+\mathbf{u})\mathrm{d}\Omega_{\mathbf{u}},
\end{equation}
and the function $N_e(\rv,u)$, 
\beq
\label{eq:ne}
N_e(\rv,u)=\int_0^u 4 \pi x^2 \,\tilde{\rho}(\rv,x)\, {\rm d} x.
\eeq
These functions have been studied and efficiently implemented by Ernzerhof and co-workers.\cite{AntZhoErn-PRA-14,ZhoBahErn-JCP-15,BahZhoErn-JCP-16} 
We now want to find a physical approximation for the crucial quantities
\beq\label{eq:nu}
\nu_i^\lambda(\rv)=N_e(\rv,R_i^\lambda(\rv)),\qquad i=2,\dots,N
\eeq
which give the expected number of electrons in a sphere of radius  $R_i^\lambda(\rv)$ centered at the reference electron in $\rv$.
We notice at this point that in an inhomogeneous system, even one-dimensional, it is not possible to write the exact SCE radii explicitly in terms of the function $N_e(\rv,u)$ obtained by spherically averaging the density around a reference electron in $\rv$ as in Eq.~\eqref{eq:spher_rho}. An exception is an homogeneous 1D system, in which $\nu_2^{\rm SCE}=\nu_3^{\rm SCE}=2$, $\nu_4^{\rm SCE}=\nu_5^{\rm SCE}=4$, etc. 

At the physical interaction strength $\lambda=1$, we expect a situation in which this extreme correlation is reduced, with all the $\nu_i$ close to $i-1$.
To illustrate this fact, we consider first an $N=2$ system, for which we have only one radius, $R_2^\lambda$, which from Eq.~\eqref{eq:edens_mod} will be equal to
\beq
R_2^\lambda(\rv)= \frac{1}{v_H(\rv)+2 w_\lambda (\rv)}, 
\label{eq:rad_inver}
\eeq
showing that, for $N=2$,  $R_2^\lambda(\rv)$ is the screening length associated with the Hartree-exchange-correlation potential when its response part is removed.\cite{LeeGriBae-ZPA-95,GriBae-PRA-96,GriMenBae-JCP-16}
Given that for two-electron systems highly accurate $w_\lambda(\rv)$ have been computed,\cite{IroTea-MP-15,locpaper} we can use  Eq.~\eqref{eq:rad_inver} to obtain the `exact' $R_2^\lambda(\rv)$.  In Fig.~\ref{fig_invrad} we show the corresponding $\nu_2^{\lambda}(\rv)$ for the hydride ion at $\lambda=0$, $\lambda=1$, $\lambda \to \infty$. We clearly see that in the physical system  $\nu_2^{\lambda=1}(\rv)$ is much closer to 1 than in the SCE extreme case, which allows for much larger numbers $\nu_2^{\infty}(\rv)$, getting equal to 2 at the nucleus, as in the homogeneous 1D solution.

It is clear that there are several ways to define approximations for $\nu_i^{\lambda}(\rv)$, using different ingredients. Here, our aim is to show that already very simple approximations can yield rather accurate results, and we focus on the physical $\lambda=1$ case. As said, we expect that $\nu_2^{\lambda=1}(\rv)\approx 1,\; \nu_3^{\lambda=1}(\rv)\approx 2,\dots$, and we write
\beq\label{eq:nu1}
\nu_i^{1}(\rv)=i-1+\sigma_i(\rv), \qquad i=2,\dots,N,
\eeq
yielding for the radii $R_i^{\lambda=1}(\rv)$ the equations
\beq\label{eq:R1}
R_i^{1}(\rv)=N_e^{-1}(\rv,i-1+\sigma_i(\rv)), \qquad i=2,\dots,N,
\eeq
with $\sigma_i(\rv)$ being the {\it fluctuation function}, which can push away or bring closer the $i$-th electron to the reference one with respect to the expected distance $a_i(\rv)=N_e^{-1}(\rv,i-1)$. In this first model, we consider only the case in which the $i$-th electron is pushed further, because for this case we can use again the mathematical structure of the SCE functional as a guide. More general models will be explored in future works. From the SCE theory for spherically symmetric systems,\cite{SeiGorSav-PRA-07,SeiDiMGerNenGieGor-PRA-17} we know that the derivative of the radial co-motion function $f_i(r)$ at point $r$ is inversely proportional to $4\pi f_i(r)^2\rho(f_i(r))$. We thus introduce the quantity $S_i(\rv)$
\beq
\label{eq:S}
S_i(\rv)=\frac{\partial N_e(\rv,u)}{\partial u}\Big|_{u=N_e^{-1}(\rv,i-1)}=4 \pi a_i(\rv)^2 \,\tilde{\rho}(\rv,a_i(\rv)),
\eeq
which, in analogy to the SCE structure, provides information on the derivative of the $R_i^{1}(\rv)$ at $\sigma_i=0$. When $S_i(\rv)$ is small, the derivative of the $R_i(\rv)$ will be very large, and we expect the electron to be pushed further, with $\sigma_i$ approaching the average value $1/2$ (which is exactly in between two expected positions). When $S_i(\rv)$ is large, the derivative of $R_i(\rv)$ is very small and we expect it to stay close to $\sigma_i=0$ (or even become slightly negative, a possibility not considered here).
Thus, for constructing the $\mod$ functional at the full coupling strength, hereinafter the $\mod$-1 functional, $W_1^\mod[\rho]=\int\rho(\rv)w_1^\mod(\rv)d\rv$, we use a simple gaussian ansatz
\beq \label{eq:sigma_1}
\sigma_i(\rv)=\frac{1}{2} \mathrm{e}^{-b\,S_i(\rv)^2},
\eeq
where $b=5$ has been chosen to optimize the He atom $W_1[\rho]$. Equations~\eqref{eq:edens_mod}, \eqref{eq:spher_rho}, \eqref{eq:ne} and \eqref{eq:nu1}-\eqref{eq:sigma_1} completely define $w_1^\mod(\rv)$. 
\begin{table}[]
\centering
\caption{Atomic (ionic) repulsion energies $W_{1}[\rho]$ obtained by the MRF-1 model and PBE are compared to reference $W_1[\rho]$, obtained with \textsc{Gamess-US} package \cite{GAMESS} using full-CI (for the first four systems) and CCSD wavefunctions (other systems). The aug-cc-pCVXZ basis set of Dunning\cite{Dun-JCP-89} has been used ($X=6$ for He and H$^-$, $X=5$ for F$^-$ and Ne, $X=T$ for Be and Li$^-$ and $X=Q$ for the other atoms). The SCE values $W_\infty[\rho]$ computed from the same densities are also reported. }
\label{tab_atoms}
\begin{tabular}{@{}lllll@{}}
\toprule
atom/ion & Reference              & ~~MRF-1     & ~~PBE     & ~~SCE                    \\ \midrule \hline
He       & -1.1029                & -1.1844 & -1.1047 & -1.4982                \\ 
H$^-$    & -0.4532                & -0.4681 & -0.4413 & -0.5689                \\
Be       & -2.8341                & -2.8044 & -2.8430  & -4.0195                \\
Li$^-$   & -1.9462                & -2.1170  & -1.9617 & -2.7308                \\
F$^-$    & -10.889                & -10.741 & -10.997 & -16.940                 \\
Ne       & -12.765                & -12.823 & -12.876 & -20.041                \\
Mg       & -16.701                & -16.365 & -16.913 & -26.709                \\
Cl$^-$   & -28.89                 & -28.48  & -29.19  & -47.26                 \\
Ar       & -31.35                 & -31.19  & -31.68  & -51.49                 \\
Ca       & -35.60                  & -35.92  & -36.85  & -60.34                 \\
MAE      &~~~~- & ~~0.17    & ~~0.24    & ~~~~- \\ \bottomrule \hline
\end{tabular}
\end{table}

\begin{figure}
\includegraphics[width=8.5cm]{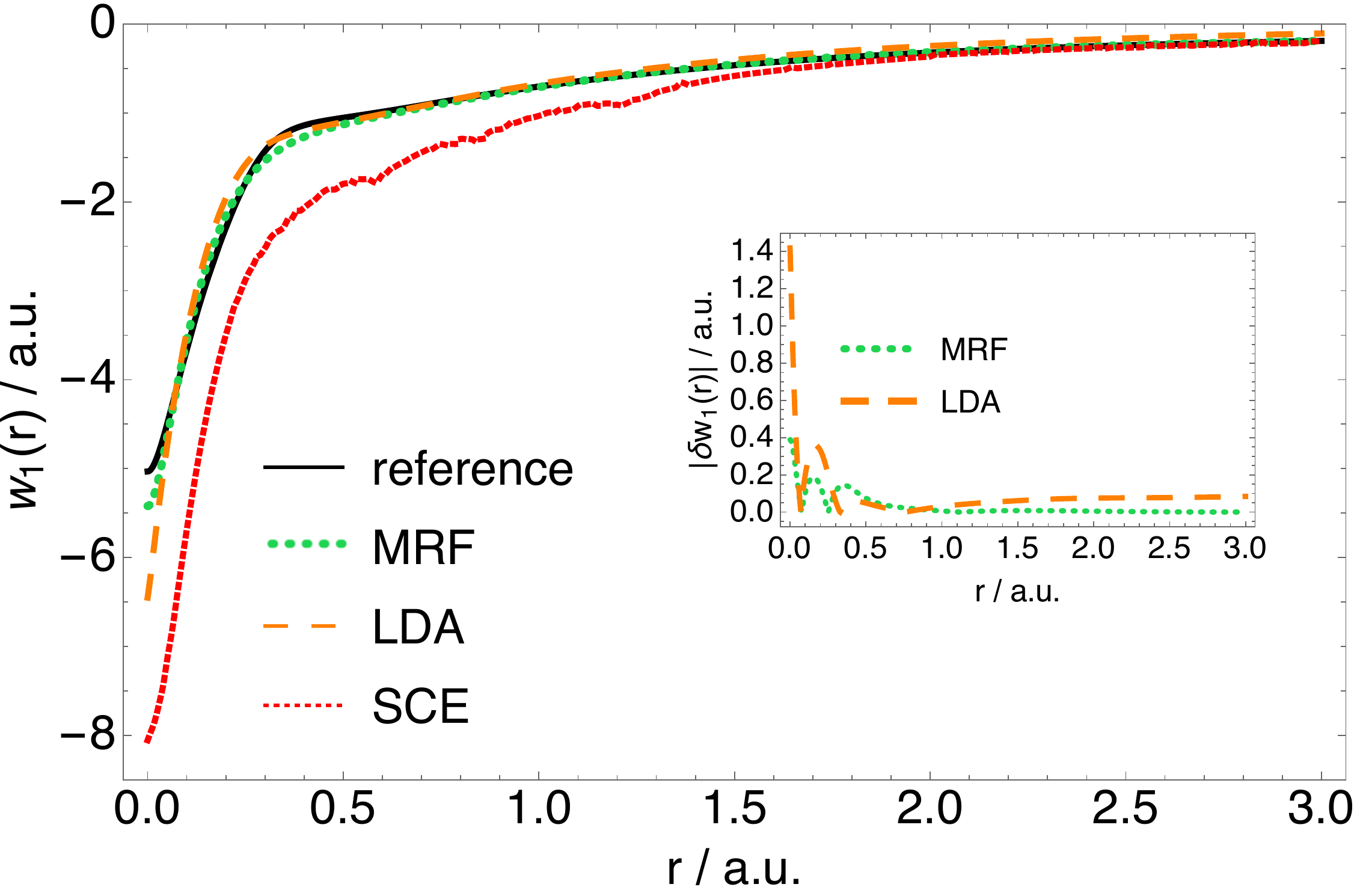}
\includegraphics[width=8.5cm]{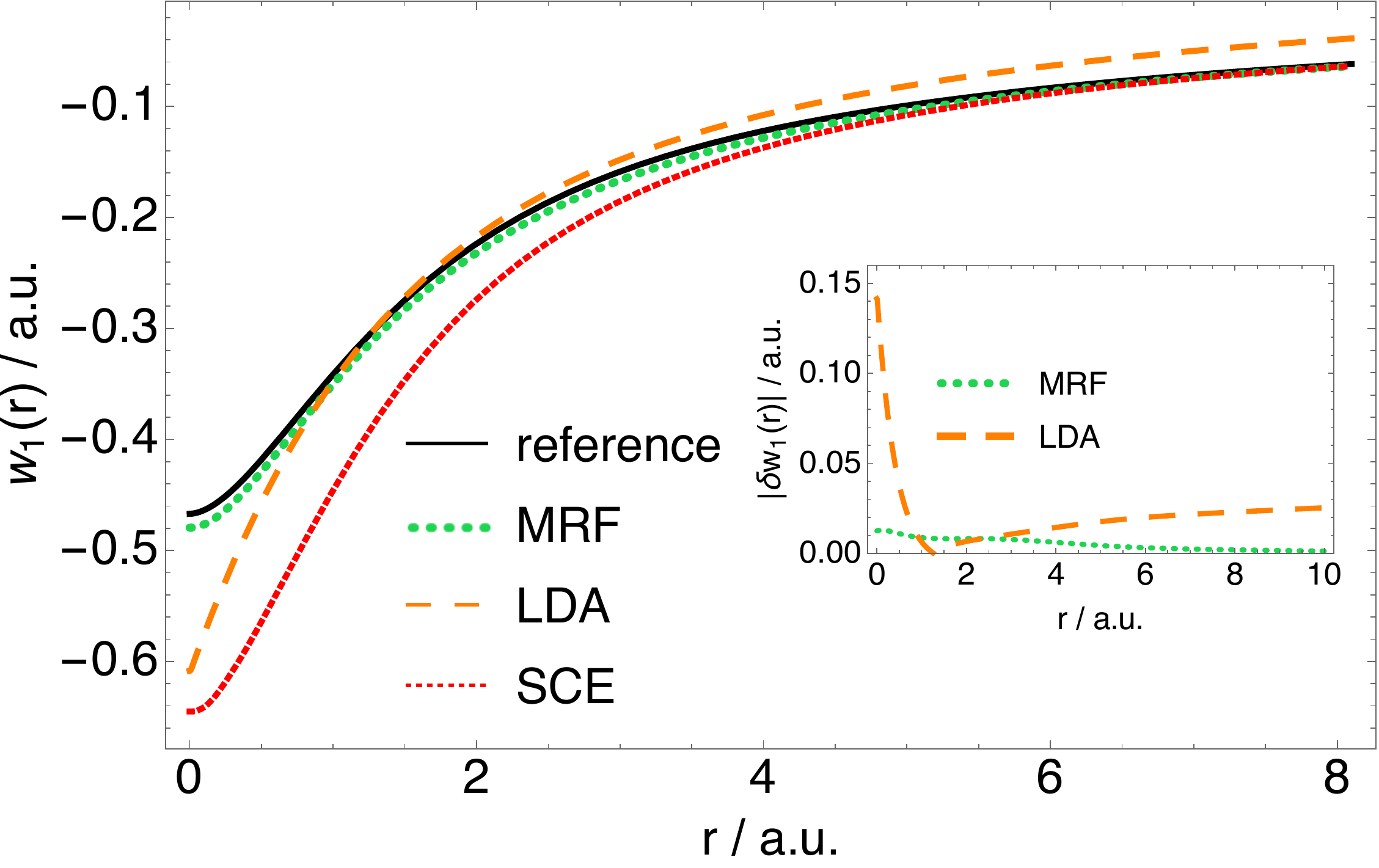}
\caption{Energy densities at full coupling strength $w_1(r)$ as a function of distance from the nucleus, $r$/a.u., obtained from the present model (MRF), from the local-density approximation (LDA),\cite{PerWan-PRB-92} and from the strictly-correlated electrons functional (SCE), all evaluated on accurate densities, for Ne (upper panel) and H$^-$ (lower panel). The reference $w_1(r)$ are obtained at the full-CI and CCSD level of theory, as in Refs.~\onlinecite{MirSeiGor-JCTC-12,locpaper}, by using the aug-cc-pCVTZ and aug-cc-pV6Z basis sets\cite{Dun-JCP-89} for Ne and H$^-$, respectively. Insets show the absolute error of approximate energy densities, $\delta w_{1}(r)=w_{1}-w_{1}^{\rm apx}(r)$.}
\label{fig_edens}
\end{figure} 
In Table~\ref{tab_atoms} we compare $W_1[\rho]$ obtained with the MRF-1 model with corresponding reference values (full-CI/CCSD), PBE and SCE ones ($W_\infty[\rho]$) for several closed-shell atomic (ionic) systems evaluated on accurate (CCSD) densities. The PBE values have been obtained by using the scaling relation\cite{LevPer-PRA-85,LevPer-PRB-93}, with $\rho_\gamma(\rv)=\gamma^3\rho(\gamma\,\rv)$,
\beq
W_1^{\rm DFA}[\rho]=E_x^{\rm DFA}[\rho]+2 E_c^{\rm DFA}[\rho]-\frac{\partial E_c^{\rm DFA}[\rho_\gamma]}{\partial \gamma}\Big|_{\gamma =1}.
\eeq
From Table~\ref{tab_atoms} we can see that our model, even with the very simple ansatz for $\sigma_i(\rv)$ of Eq.~\eqref{eq:sigma_1}, gives repulsion energy much closer to the physical ones with respect to SCE. Their quality is comparable to that of PBE, with MAE somewhat smaller (0.17~a.u.~vs.~0.24~a.u.). The purpose here is not to reach high accuracy (which requires optimization and further studies of the $R_i$), but to show that functional approximations based on modeling the quantity $\sigma_i(\rv)$ is a very promising strategy, because already a primitive non-optimized model performs very well. Even more interesting than the global $W_1^\mod[\rho]$ values are the energy densities:
in Fig.~\ref{fig_edens} we compare $w_1^{\rm MRF}(\rv)$ with the reference $w_1(\rv)$ for the neon atom (top panel) and the hydride ion (bottom panel). We also show $w_{1}(\rv)$ obtained with the LDA functional from the PW92 parametrisation,\cite{PerWan-PRB-92,Mar-PR-58} 
$w_{1}^{\rm LDA}(r_s)= \frac{1}{r_s}\frac{\partial}{\partial{r_s}}\big({r_s}^2 \epsilon_{xc}(r_s)\big)$.
We see that $w_1^{\rm MRF}(\rv)$ is in good agreement with the reference $w_1(\rv)$ in the case of Ne, but also in the more correlated case\cite{MirSeiGor-JCTC-12,locpaper} of H$^-$, again improving dramatically with respect to SCE. From the insets of the same figure we can see that the local error of our model is very small, vanishing for large $\rv$ due to the  correct $-\frac{1}{2 |\rv|}$ asymptotic behaviour, arising from the proper normalization of Eq.~\eqref{eq:p2}. The availability of DFAs energy densities in this gauge is rather limited, and beyond LDA it is restricted to few approximations\cite{BecRou-PRA-89,TaoBulScu-arx-16} to the exchange energy density ($\epsilon_x(\rv)=w_0(\rv)$). For instance, the gauge incompatibility\cite{PerRuzSunBur-JCP-14} of the generalized gradient approximation (GGA) exchange energy densities and the exact $\epsilon_x(\rv)$, which is in the gauge of Eq.~\eqref{eq:wxc_def}, has been a major hurdle for the development of local hybrid DFAs.\cite{JarScuErn-JCP-03,ArbKau-CPL-07} 

\begin{figure}
\includegraphics[width=8.5cm]{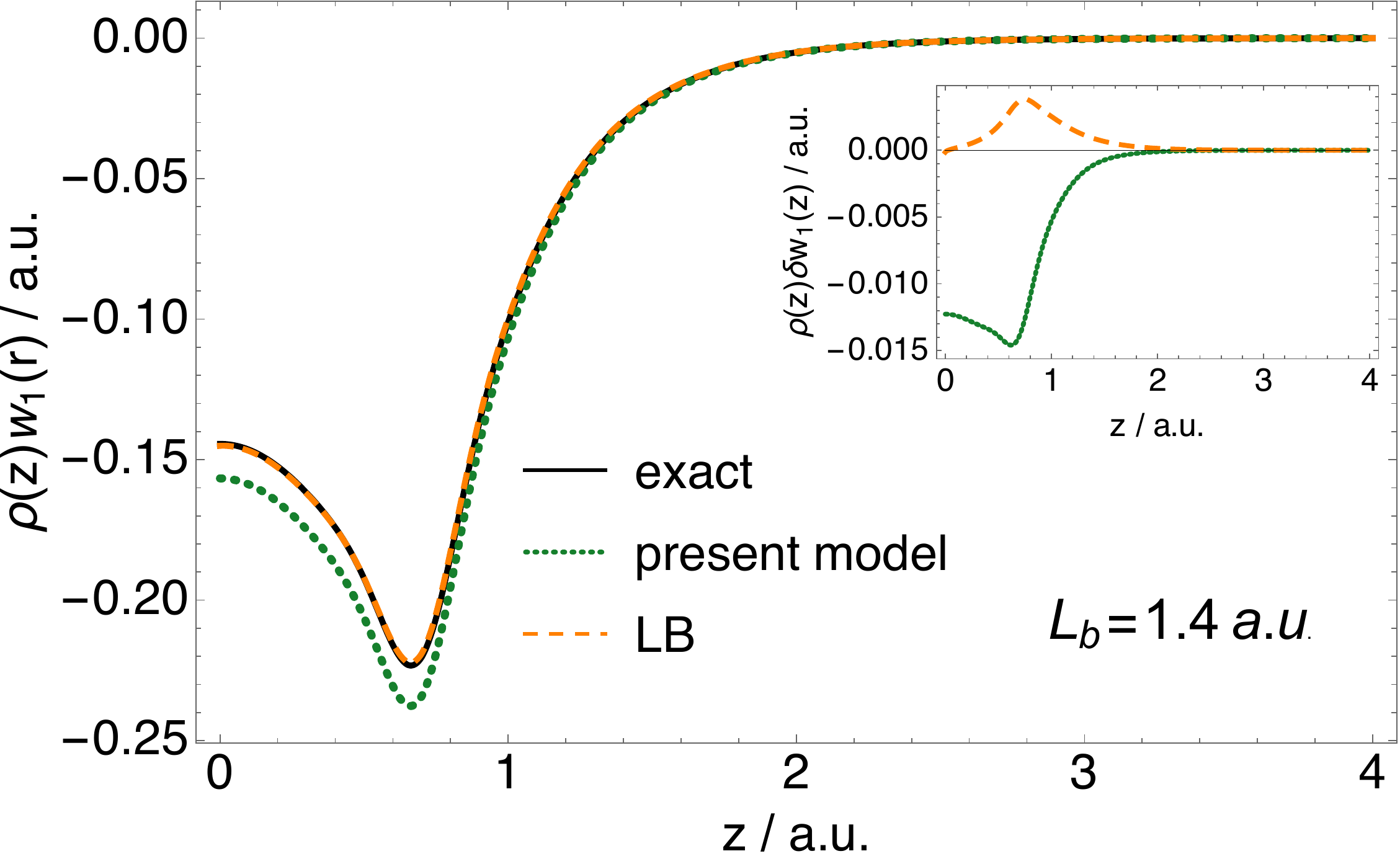}
\includegraphics[width=8.5cm]{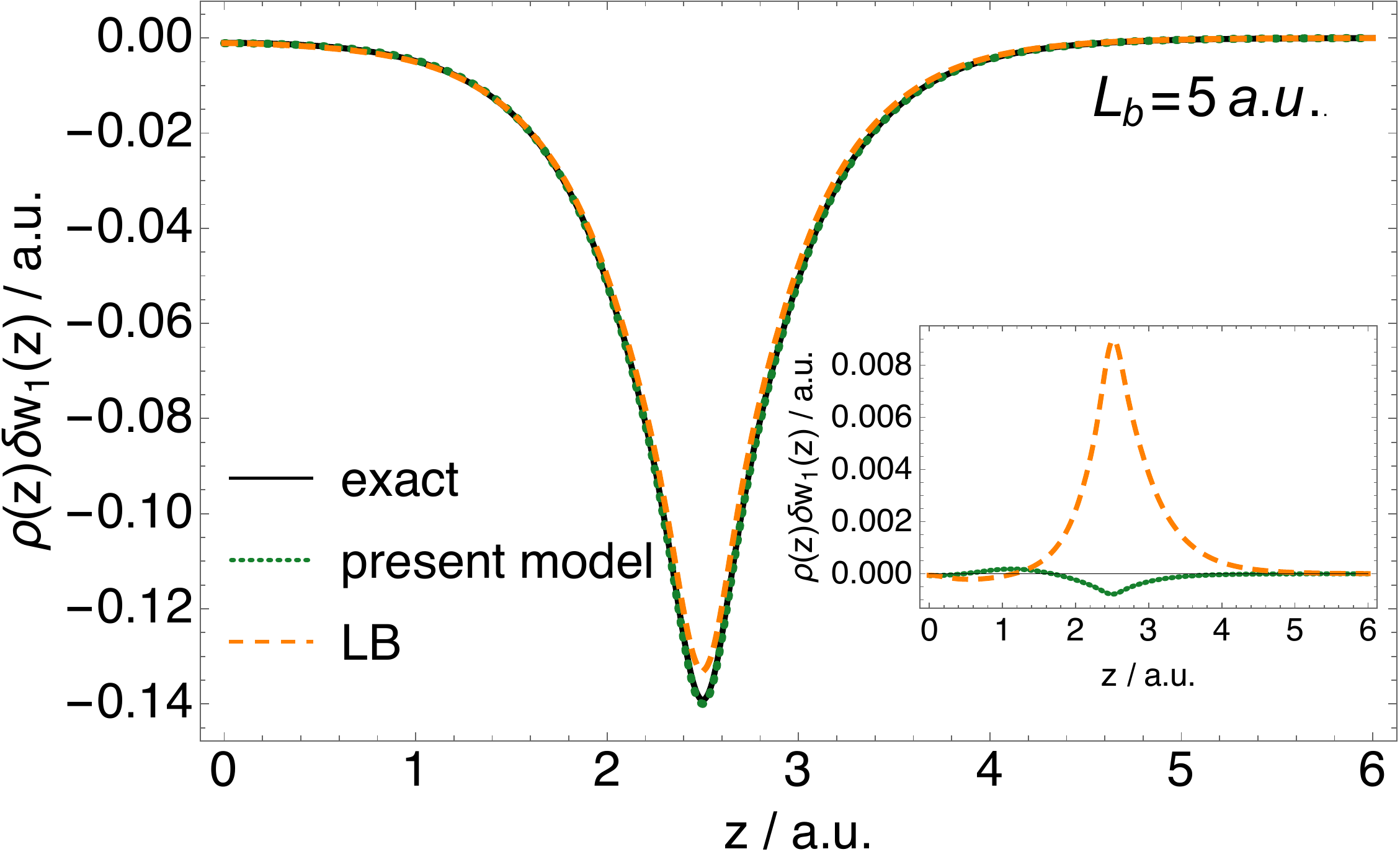}
\includegraphics[width=8.5cm]{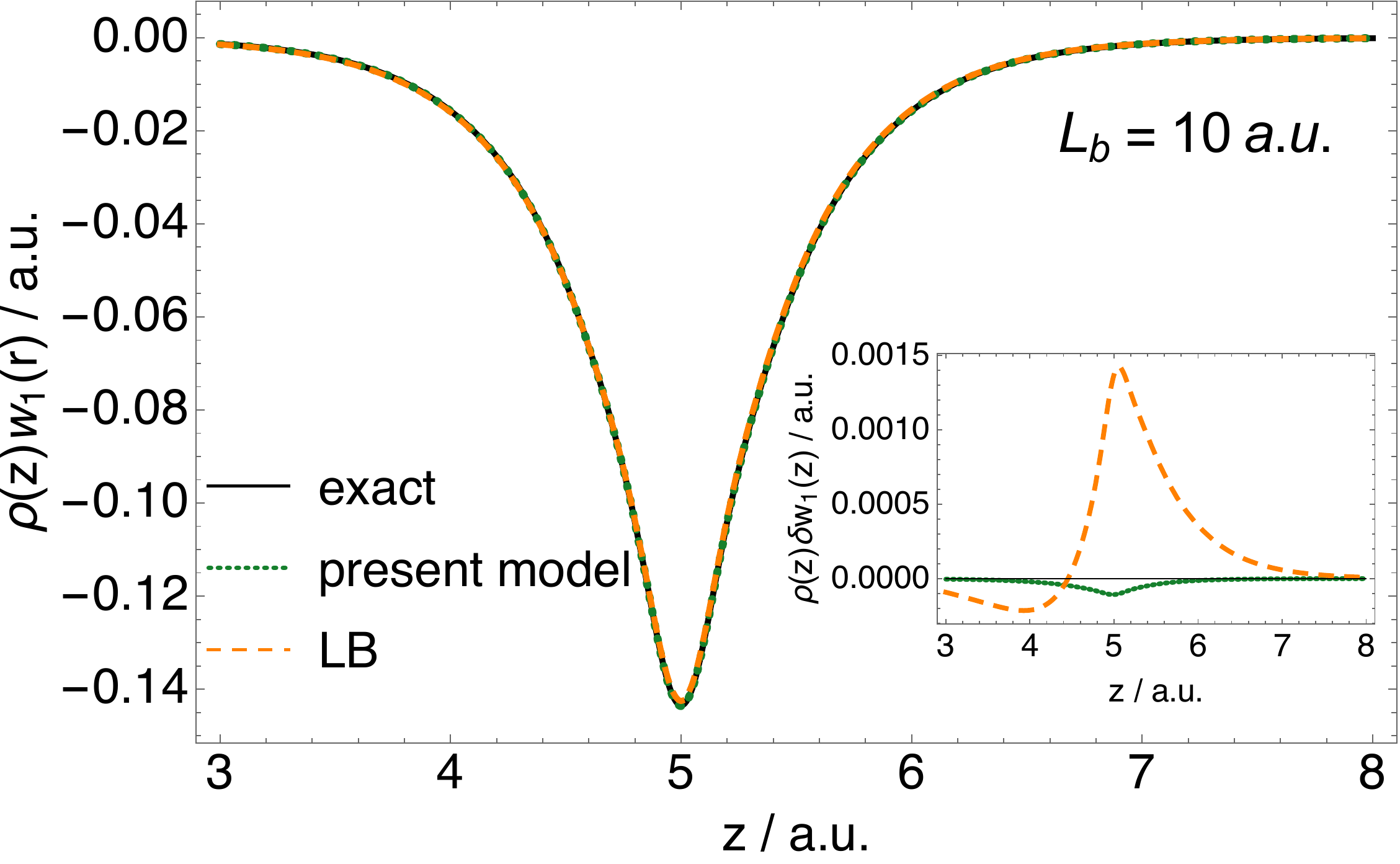}
\caption{Energy densities at full coupling strength as a function of the distance from the bond midpoint $z$ along the internuclear axis for the H$_2$ molecule at different bond-lengths $L_b$ obtained with MRF-1 using accurate FCI/aug-cc-pCVTZ densities. Reference energy densities and those obtained with the LB local interpolation scheme are from Ref.~\onlinecite{locpaper}. }
\label{fig_edensh2}
\end{figure}

The main point of introducing the full non-local dependence is of course to treat static and strong correlation. In Fig.~\ref{fig_edensh2} we show the energy density $w_1^{\rm MRF}(\rv)$ for the H$_2$ molecule at different bond lengths $L_b$ along the internuclear axis, compared with accurate ones from Ref.~\onlinecite{locpaper}. For comparison, we also show $w_1(\rv)$ obtained from the interpolation model of Liu and Burke (LB)\cite{LiuBur-PRA-09} applied to energy densities, with the exact $w_0(\rv)$, $w'_0(\rv)$ and $w_\infty(\rv)$ as input ingredients.\cite{locpaper} As we can see from Fig.~\ref{fig_edensh2}, in the equilibrium region the MRF-1 energy densities are still somewhat lower than the reference ones, whereas the LB is highly accurate, as it is also the case with atoms.\cite{locpaper,VucIroWagTeaGor-PCCP-17} However, we can also see that in the stretched case ($L_b=10$~a.u.) the MRF-1 energy densities are very accurate, even more accurate than the LB interpolated ones (note again that they use the exact $w_0(\rv)$, $w'_0(\rv)$ and $w_\infty(\rv)$ as input for the interpolation, whose error is already small).  While in the stretched H$_2$ molecule the static correlation effects are dominant, at intermediate bond lengths, around $L_b\sim 5.0$~a.u., there is a subtle interplay between dynamic and static correlation effects.\cite{VucIroWagTeaGor-PCCP-17} This region can be even more challenging for DFAs than the stretched case, given that certain DFAs which dissociate H$_2$ correctly fail in this scenario yielding a positive ``bump'' (see, e.g., Refs~\onlinecite{FucNiqGonBur-JCP-05,PeaMilTeaToz-JCP-08,locpaper,ZhaRinPerSch-PRL-16}). We can see that the MRF-1 energy densities are very accurate at $L_b=5.0$~a.u., hardly distinguishable from the reference ones. 

\begin{figure}
\includegraphics[width=8.5cm]{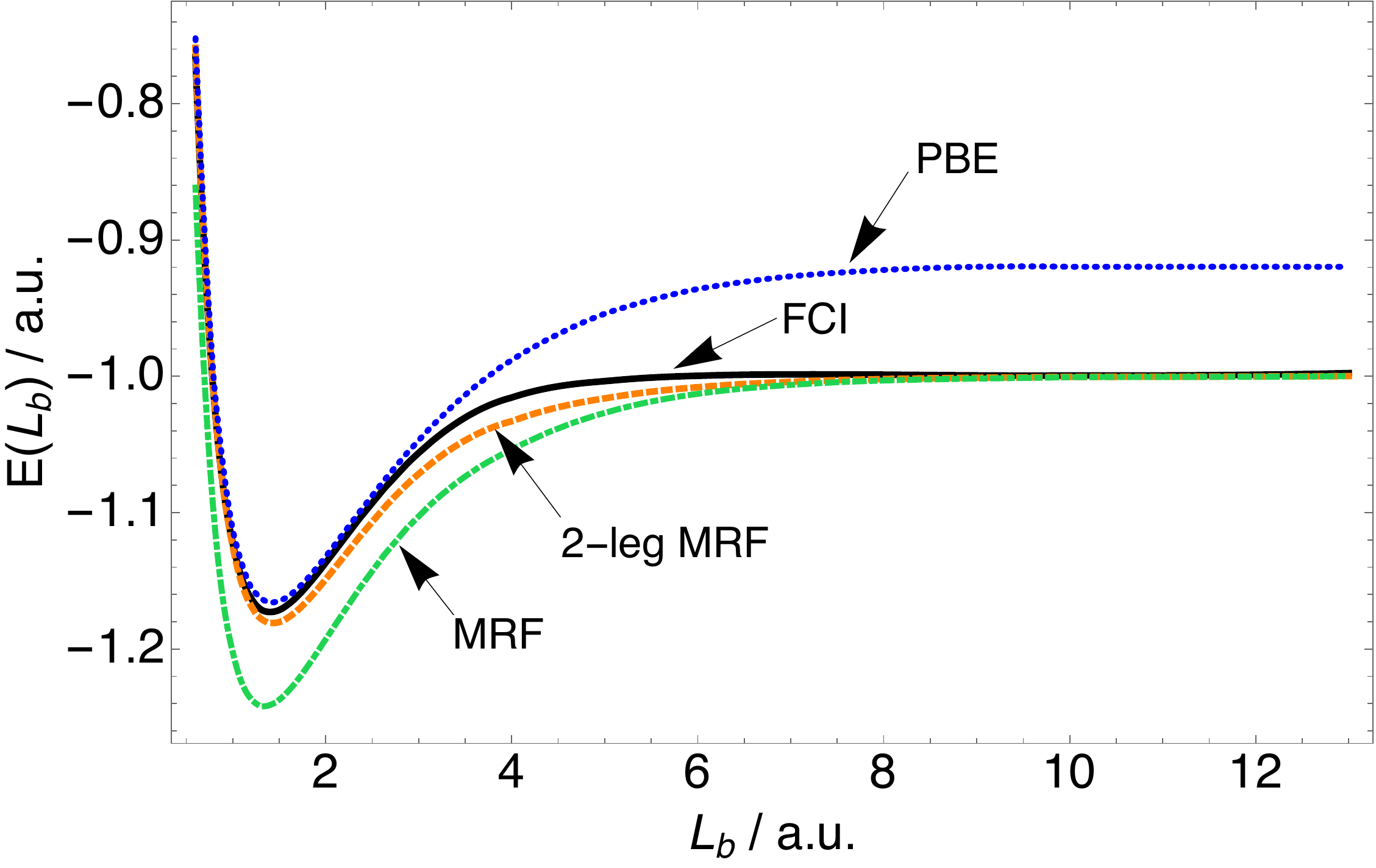}
\caption{The H$_2$ molecule dissociation curve as a function of the internuclear distance $L_b$/a.u. obtained with the MRF-1 and 2-leg MRF approaches presented in this work, compared to restricted PBE and FCI. All the curves have been obtained using the aug-cc-pCVTZ basis set\cite{Dun-JCP-89}}
\label{fig_h2cur}
\end{figure}

In Fig.~\ref{fig_h2cur} we show the dissociation curve for the H$_2$ molecule obtained using the MRF-1 functional evaluated on the accurate FCI/aug-cc-pCVTZ densities. We can see that around equilibrium MRF-1 underestimates the total energy, because it misses the positive kinetic correlation component $T_c[\rho]$ and slightly underestimates the exact $W_1[\rho]$, as already shown in Fig.~\ref{fig_edensh2}. Despite missing $T_c[\rho]$, MRF-1 dissociates H$_2$ correctly, because $T_c[\rho]$ vanishes as the H$_2$ dissociates into atoms. To recover the missing $T_c[\rho]$ component, one can combine $W_1^{\rm MRF}[\rho]$ with the quantities from the weak coupling limit, namely $W_0[\rho]$ and $W'_0[\rho]$, to interpolate $W_\lambda[\rho]$ and thus obtain $E_{\rm xc}[\rho]$. For this purpose, we employ a very simple interpolation form, the two-legged representation,\cite{BurErnPer-CPL-97,locpaper,VucIroWagTeaGor-PCCP-17} which has recently been used to construct a tight lower bound to correlation energies.\cite{VucIroWagTeaGor-PCCP-17} This form reads as
\begin{subequations}\begin{align}
W_{\lambda}[\rho] &= 
\begin{cases}
W_{0}[\rho] + \lambda W'_{0}[\rho], & \lambda \leqslant X_{c} \\
W_{1}[\rho], & \lambda > X_{c}
\end{cases} \label{eq:2leg} \\[2ex]
X_{c} &= \frac{W_{1}[\rho] - W_{0}[\rho]}{W'_{0}[\rho]}. \label{eq:xlam}
\end{align}\end{subequations}
As in this work we use $W_1^{\rm MRF}[\rho]$ as an approximation to $W_{1}[\rho]$, we call this approach the ``2-leg MRF'', and from Fig.~\ref{fig_h2cur} we can see that it substantially improves the MRF-1 energies. In this case very similar results are obtained if we do the interpolation on the energy densities rather than on integrated quantities. Besides dissociating correctly H$_2$, the dissociation of H$_2^+$ is also correctly described within the MRF-1 and 2-leg MRF approaches, because our model of Eq.~\eqref{eq:p2} is equal to $0$ for all $N=1$ systems.  

Finally, one may wonder if the MRF would encounter problems for extended systems. As a paradigmatic example, we consider the uniform electron gas (UEG) with density $\rho=(\frac{4}{3}\pi r_s^3)^{-1}$, for which $N_e(\rv,u)=N_e(u)=u^3/r_s^3$ and $N_e^{-1}(i-1)=r_s(i-1)^{1/3}$. Then, by using the model pair density of Eq.~\eqref{eq:p2} we obtain $w_1^\mod(r_s)=\tilde{w}(r_s)/r_s$ with
\beq
\label{eq:MRFUEGcon}
\tilde{w}(r_s)=\lim_{N\to\infty}\frac{1}{2}\left[ \sum_{i=2}^N\frac{1}{(i-1+\sigma_i(r_s))^{1/3}}-\frac{3}{2}N^{2/3}\right],
\eeq
where $\sigma_i(r_s)$ is given by Eq.~\eqref{eq:sigma_1} and, from Eq.~\eqref{eq:S}, $S_i=3(i-1)^{2/3}/r_s$. This expression has a fast $N\to\infty$ convergence and when $\sigma_i=0$ can be evaluated in closed form. It yields reasonable values for the UEG, with a maximum relative error of 23\%. The function $\tilde{w}(r_s)$ of Eq.~\eqref{eq:MRFUEGcon} has the same qualitative behavior of the exact one: it is monotonically decreasing with $r_s$, bounded between the two limting values $\tilde{w}(0)=-0.487$ and $\tilde{w}(\infty)=-0.7564$, not so far (considering the simplicity of the model for $\sigma_i(r_s)$) from the exact ones, $-\frac{3}{4}(\frac{3}{2\pi})^{2/3}\approx-0.458$ and $\approx -0.876$, respectively (this latter value is currently a matter of discussion, see Refs.~\onlinecite{LewLie-PRA-15,SeiVucGor-MP-16}). The UEG also illustrates the physics of the model: when $r_s\to 0$ (weak correlation) $\sigma_{i\geq 2}\to 0$, while when $r_s\to \infty$ (strong correlation), $\sigma_i\neq 0$ for $i$ larger and larger (long-range fluctuations become more and more important). It also suggests that for extended systems the explicit functional can be confined to a set $i< i_{\rm max}$, and the rest can be resummed. The value $i_{\rm max}$ is determined by correlation (for example, in the UEG it is automatically determined by $\sigma_i(r_s)$).

In summary, we have proposed a strategy to build fully non-local DFAs inspired by the mathematical structure of the exact XC functional in the strong coupling limit, reducing the problem to the construction of the fluctation function $\sigma_i(\rv)$ in terms of $S_i(\rv)$ of Eq.~\eqref{eq:S}. Already an extremely simple model such as the one of Eq.~\eqref{eq:sigma_1} is {\it locally} accurate, it is able to dissociate correctly the H$_2$ and H$_2^+$ molecules, and gives very  reasonable results for the uniform electron gas. 
We thus believe that the nonlocal structure of our functional, which goes beyond the Jacob's ladder framework, opens up new perspectives for the development of XC functionals able to tackle strong correlation. Although the functional is highly nonlocal, it can be obtained at a computational cost comparable to that of the NLR and shell functionals, which have been recently implemented in a very efficient way.\cite{BahZhoErn-JCP-16} 
Many strategies to improve the accuracy can be pursued: the inclusion of kinetic correlation through interpolation along the adiabatic connection (as in Fig.~\ref{fig_h2cur}); trying to model directly the $\lambda$-dependence of $\sigma_i$;  the generalisation to non-integer number of electrons\cite{MirSeiGor-PRL-13} and spin densities; improving the accuracy for the UEG, and adding the dependence on the gradient of $S_i(\rv)$.
The functional can also be readily applied to  other dimensionalities, e.g. electrons confined in quasi-1D and quasi-2D geometries, for which the SCE approach has already proven very useful.\cite{MalMirCreReiGor-PRB-13,MenMalGor-PRB-14} It can be also applied to other isotropic interactions, such as the error function used in range separation\cite{Sav-INC-96} but also effective interactions for ultracold quantum gases.\cite{MalMirMenBjeKarReiGor-PRL-15}

This work was supported by the Netherlands Organization for Scientific Research (NWO) through an ECHO grant (717.013.004) and the European Research Council under H2020/ERC Consolidator Grant corr-DFT (Grant No. 648932).

\bibliography{biblioPaola,biblio_spec,biblio1,biblio_add,biblio_add2}

\end{document}